\documentstyle[svcon,psfig]{report}
\def\about{$\sim$}

\def\approxlt{\ifmmode \rlap{$<$}{}_{{}_{{}_{\textstyle\sim}}} \else%
$\rlap{$<$}{}_{{}_{{}_{\textstyle\sim}}}$\fi}
\begin{document}
\pagenumbering{arabic}
\chapter{Quantifying Rapid Variability in Accreting Compact Objects}
\chapterauthors{M. van der Klis}
\begin{abstract}
I discuss some practical aspects of the analysis of millisecond time
variability X-ray data obtained from accreting neutron stars and black
holes. First I give an account of the statistical methods that are at present
commonly applied in this field. These are mostly based on Fourier
techniques. To a large extent these methods work well: they give astronomers
the answers they need. Then I discuss a number of statistical questions that
astronomers don't really know how to solve properly and that statisticians may
have ideas about. These questions have to do with the highest and the lowest
frequency ranges accessible in the Fourier analysis: how do you determine the
shortest time scale present in the variability, how do you measure steep
low-frequency noise. The point is stressed that in order for any method that
resolves these issues to become popular, it is necessary to retain the
capabilities the current methods already have in quantifying the complex,
concurrent variability processes characteristic of accreting neutron stars and
black holes.
\end{abstract}

\section{Introduction}\label{intro}

The purpose of this talk is to explain to statisticians how astrophysicists,
mostly using Fourier transform techniques, go about analyzing X-ray
time-series data obtained from accreting compact objects (neutron stars and
black holes), and to point out a few problems with the usual approaches. The
point will be made, that the conglomerate of statistical methods that is being
applied in this branch of high-energy astrophysics, even though most
definitely not always rigorous, on the whole serves its purpose well and is
providing astronomers with the quantitative answers they require. This talk
will aim, however, at a few areas where we run into problems, and where more
statistical expertise might help. In thinking about how to solve the problems
that I shall outline, it will be important to keep an eye on what the
capabilities of the existing methods are, as those capabilities will need to
be preserved in whatever new approach one would like to propose.

\begin{figure}[hbt]
\begin{center}
\begin{tabular}{c}
\psfig{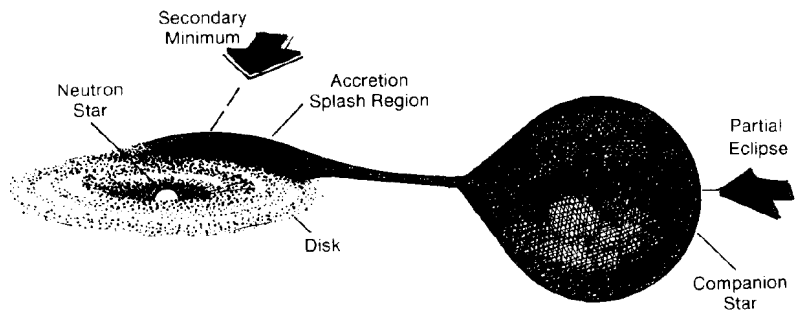}
\end{tabular}
\caption{X-ray binary star.\label{lmxb}}
\end{center}
\end{figure}

Mostly, accreting neutron stars and black holes occur in double star systems
known as X-ray binary stars, where a normal star and the compact object are in
a close orbit around each other (Fig.\,\ref{lmxb}). Matter flows from the normal star to
the compact object by way of a flat, spiraling flow called an accretion disk,
and finally accretes onto the compact object. A large amount of energy is
released in this process (typically 10$^{36}$ to 10$^{38}$\,erg/sec), and is
emitted, mostly in the form of X-rays. The characteristic variability
timescale for the X-ray emitting regions is predicted (and observed) to be
very short (less than a millisecond). By studying the properties of this rapid
X-ray variability it is possible to extract a great deal of information about
the flow of matter onto the compact object, and, indirectly, about the object
itself. See van der Klis (1995) for a recent review of the results of studies
of this type.

\section{The data}\label{data}

In order to understand the character of the date we are dealing with, I shall
follow the flow of information from star to computer. An X-ray binary star
emits X-ray photons at a very large rate (say, 10$^{46}$\,photons/sec). For
all practical purposes, the X-ray photon rate produced by the star can be
considered as a continuous function of time $I(t)$. These photons are emitted
isotropically, or at least over a solid angle of order 4$\pi$\,sterad. Because
the X-ray detector onboard the X-ray satellite spans only a very small solid
angle as seen from the X-ray star, only a very small fraction $\epsilon$ (say,
10$^{-43}$) of the photons is detected in the instrument. The time series of
photon arrival times $t_i, i = 1,\dots,N_{phtot}$ (with $N_{phtot}$ the total
number of detected photons) is the information that, ideally, we would like to
have available for analysis. However, typically, instrumental limitations (and
the maximum telemetry rate) prevent the registration and transmission of all
these arrival times. Therefore, onboard the satellite, the data are {\it
binned} into equidistant time bins. The information that is finally
telemetered to the ground station consists of a sequence of numbers $x_m, m =
0,\dots,N_{tot}-1$, where $N_{tot}$ is the total number of time bins, $x_m$
representing the number of photons that was detected during time bin $m$. In
most cases, and unlike the usual case in astronomy, the time bins are
equidistant and contiguous (no gaps). The statistical problem facing us can be
summarized as follows: ``{\it Given $x_m, m=0,\dots,N_{tot}-1$, reconstruct as
much as possible about $I(t)$}''.

We shall assume that for the bright X-ray binaries that I am discussing here,
the rate of background photons can be considered to be negligible, and that
there are no other relevant effects affecting the photon arrival time series
than the huge geometrical dilution factor just described.\footnote{This is of
course not exactly true. The background does not pose large problems in
practice, as it just constitutes an additional source of detections (of
photons as well as charged particles) that is not strongly variable on the
time scales we are interested in and uncorrelated to the fluctuations due to
the star. Detector deadtime processes leading to ``missed'' photons {\it do}
constitute a serious complication (e.g., van der Klis 1989, Mitsuda and Dotani
1989, Vikhlinin et al. 1994, W. Zhang 1995) that will be ignored here.}
Therefore, if the X-ray star does not vary intrinsically, the $t_i$ will to a
high degree of precision be uniformly and randomly distributed, so that the
$x_k$ follow Poisson statistics appropriate to a rate $\mu =
\langle x_m\rangle$, with a standard deviation $\sigma_{x_m}$ equal to
$\sqrt\mu$.

Of course these stars {\it do} vary. The variability time scales of interest
are $\approxlt$1\,millisecond, and the detected photon rates
10$^2$--10$^5$\,photons/sec, which means that the time bins must be chosen
such that typically there are on average only a few (sometimes $\ll$1) photons
per time bin, and $\sigma_{x_m}$ is of order $\mu$ (sometimes
$\sigma_{x_m}\gg\mu$). As the intrinsic variability of the star often has a
relatively small amplitude, only a few percent of the total flux, it is clear
that we are in a low signal-to-noise regime (with the ``signal'' the intrinsic
stellar variability and the ``noise'' the Poisson fluctuations). Fortunately,
typical observations span 10$^3$ to 10$^5$\,sec, so that the number of time
bins $N_{tot}$ (the number of ``measurements'') is 10$^6$ to 10$^8$, which
allows us to recover the signal from the noise. The techniques used for this
are described in the next section.

\section{Standard analysis}\label{ana}

The standard approach (see van der Klis 1989) is to divide the time series
into $M$ equal-length segments of $N$ time bins $x_k, k=0,\dots,N-1$ each (so,
ideally $N_{tot}=MN$), to calculate the discrete Fourier transform of each segment:
$$ a_j \equiv \sum_{k=0}^{N-1}x_k e^{2\pi ijk/N} \qquad j = 0,\dots,N/2, $$
to convert this into a power spectrum for each segment
$$ P_j \equiv {2\over a_0}|a_j|^2 \qquad j = 0,\dots,N/2, $$
and then to {\it average} these power spectra (see below). Note that in our
application $a_0\equiv\sum_{k=0}^{N-1}x_k=N_{ph}$, the number of photons
detected in the segment. With this power spectral normalization, due to Leahy
et al. (1983), it is true that if the $x_k$ are distributed according to the
Poisson distribution, then the $P_j$ follow the $\chi^2$ distribution with 2
degrees of freedom, so $\langle P_j\rangle=2$ and $\sigma_{P_j}=2$.

This white noise component with mean level 2 and standard deviation 2, induced
in the power spectrum due to the Poisson fluctuations in the time series, is
called the ``Poisson level''. It can be considered as ``background'' in the
power spectrum, against which we are trying to observe the other power
spectral features, which are caused by the intrinsic variability of the X-ray
binary. The physical dimension of the thus defined powers is the same as that
of the time series: $[P_j]=[a_j]=[x_k]$. Often the Y-axis of plots of power
spectra in this normalization is just labeled ``POWER'', which reflects the
fact that the physical interpretation of the $P_j$ in terms of properties of
the star is inconvenient. For this reason, in recent years another power
spectral normalization has become popular where the powers are reported as
$Q_j\equiv P_j/\lambda$, with $\lambda$ the ``count rate'', the number of
detected photons per second: $\lambda=N_{ph}/T$, where $T$ is the duration of
a segment. The $Q_j$ are dimensionless, and can be interpreted as estimates of
the power density $Q(\nu_j)$ near the frequency $\nu_j\equiv j/T$, where
$Q(\nu)$ is a function of frequency whose integral gives the fractional
root-mean-square amplitude $r$ of the variability. This latter quantity is
defined as
$$ r \equiv {\sqrt{{1\over N}\sum_{k=0}^{N-1}(x_k-\overline{x})^2} \over
\overline{x}} \qquad \hbox{\rm where} \qquad \overline{x} \equiv 
{1\over N}\sum_{k=0}^{N-1}x_k.$$
(This follows directly from Parseval's theorem.) The fractional rms amplitude
$r_{12}$ due to fluctuations in a given frequency range $(\nu_1,\nu_2)$ is
given by
$$r_{12} = \int_{\nu_1}^{\nu_2}Q(\nu)d\nu.$$
So, the physical interpretation of $Q(\nu)$ is easy: it is the function whose
integral gives you the square of the fractional rms amplitude of the
variability in the original time series. The physical unit used for $Q(\nu)$
is (rms/mean)$^2$/Hz, where ``rms'' and ``mean'' both refer to the time
series; ``rms/mean'' is just the dimensionless quantity $r$.

\begin{figure}[hbt]
\begin{center}
\begin{tabular}{c}
\psfig{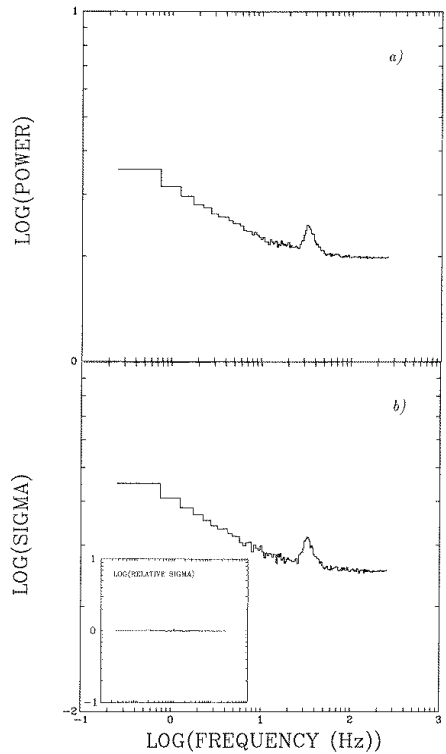}
\end{tabular}
\caption{Top: An average of 6166 power spectra of EXOSAT ME data on the X-ray
binary GX\,5$-$1 showing power-law noise and QPO. Bottom: The standard
deviation of the 6166 power values averaged in each frequency bin. Inset: the
ratio of standard deviation to mean. The standard deviation equals the mean
power as expected for $\chi^2$ distributed powers. From Van der Klis
(1989). \label{cesme}}
\end{center}
\end{figure}

The averaging of the power spectra mentioned above is usually performed both
by averaging individual power spectra (from different segments) {\it together}
(averaging the $P_j$'s with the same $j$ from the $M$ different segments) and
by averaging powers at adjacent frequencies ($P_{j+1}$ to $P_{j+W}$, say). The
main purpose of this is of course to decrease the standard deviation of the
power estimates, which in the raw spectra is equal to the mean power. The
reason to calculate many power spectra of segments of the data rather than one
very large power spectrum of the whole data set, apart from computational
difficulties with this approach, is that in this way it is possible to study
the variations in the power spectrum as a function of time.

The final step in the analysis is to fit various functional shapes $f(\nu)$ to
the power spectra using the method of $\chi^2$ minimization that is also used
in X-ray spectroscopy (the Levenberg-Marquardt method described in Press et
al. 1992, Chapter 15). Because many individual power estimates have been
averaged in the analysis process, the central limit theorem ensures that the
uncertainties of the final power estimates $\overline Q$ are approximately
normally distributed, as required for this method to work well. A problem is
what ``uncertainties'' $\sigma_{\overline Q}$ to assign to these power
estimates. Usually $\sigma_{\overline Q}=\overline Q/\sqrt{MW}$ is assumed,
where $MW$ is the number of individual powers averaged to obtain $\overline Q$
($W$ stands for the ``width'', the number of adjacent powers averaged; $M$ for
the number of averaged power spectra). This is approximately correct, and was
experimentally verified (Fig.\,\ref{cesme}), in the case that the
dynamic range of the power spectrum is dominated by the intrinsic differences
in the mean amplitude of the star's variability as a function of frequency
rather than by the stochastic fluctuations in power. If instead the stochastic
variations dominate then this procedure for estimating the uncertainties can
lead to severe underestimation of the power, as accidentally low powers will
get high weights in the fitting procedure and vice versa. A solution to this
problem that is sometimes adopted is to estimate the uncertainty in $\overline
Q$ as $f(\overline\nu)/\sqrt{MW}$, where $f(\overline\nu)$ is the fit function
and $\overline\nu$ the frequency corresponding to $\overline Q$.

\section{Applications and results}\label{results}

The method described in the previous section works. It allows astronomers to
quickly characterize the variability properties of large amounts of data, to
study the changes in the properties of the variability as a function of time
and other source characteristics, and to measure amplitudes and characteristic
time scales of the variability. The method straightforwardly extends to
simultaneous multiple time series, for example time series obtained in
different photon energy bands. For $N_{band}$ simultaneous time series, it is
possible to calculate $N_{band}(N_{band}-1)$ different cross-spectra between
them, and to look for systematic time delays in the variability as a function
of energy. A comprehensive description of the methods used for this can be
found in the paper by Vaughan et al. (1994).

\begin{figure}[hbt]
\begin{center}
\begin{tabular}{c}
\psfig{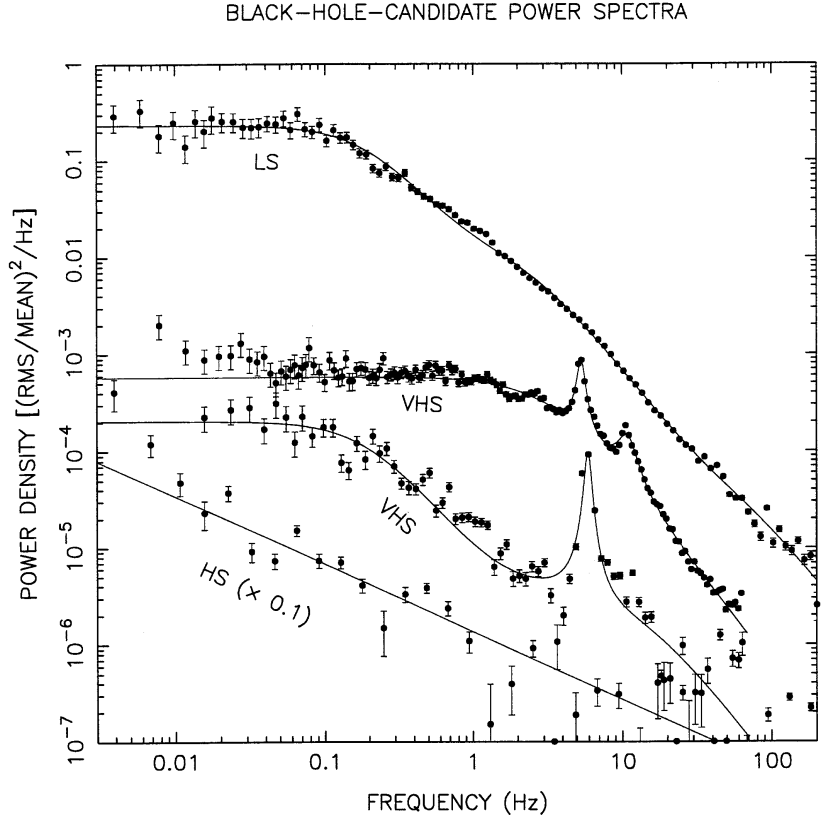}
\end{tabular}
\caption{Power spectra of black hole candidates. The spectrum labeled LS (low
state) is of Cygnus\,X-1; the other two are of GS\,1124$-$68. From Van der
Klis (1995). \label{bhcps}}
\end{center}
\end{figure}

A very important aspect is the possibility to identify different ``power
spectral components'' in the variability. This is done by studying the changes
in the shape of the power spectra as a function of time and other source
properties, such as brightness or photon energy spectrum. It usually turns out
that the simplest way to describe the changes in the power spectrum as a
function of time is in terms of the sum of a number of components whose
properties (strength, characteristic frequency) depend in a smooth, systematic
and repeatible way on, for example, brightness. If this is the case, then it
is natural to interpret these power spectral components as due to different
physical processes (or different aspects of the same process) that are all
affecting the count rates at the same time, and that have been disentangled
from each other in the analysis just described.

Examples of power spectral components that are distinguished in practice are
``power law noise'', ``band-limited noise'' and ``quasi-periodic oscillations
(QPO)''. All of these are all presumed to be stochastic processes in the time
series that cause, respectively, a component in the power spectra that fits a
function $f_{PL}(\nu) = C_{PL}\nu^{-\alpha}$, one that fits $f_{BLN}(\nu) =
C_{BLN}\nu^{-\alpha}e^{-\nu/\nu_{cut}}$ and one that fits a Lorentzian
$f_{QPO}(\nu) = C_{QPO}/(\nu-\nu_{QPO})^2+(\Gamma/2)^2)$. For a component to
be called QPO it is sufficient that the corresponding power spectral component
has a shape approximately described by $f_{QPO}(\nu)$ with the peak
full-width-at-half-maximum (FWHM) $\Gamma$ less than half its centroid
frequency $\nu_{QPO}$; of course this definition is arbitrary.

Figure\,\ref{bhcps} shows a number of actually observed power spectra, plus
the functions that were fit to them in order to describe them in terms of the
power spectral components just mentioned. QPO peaks, band-limited noise and
power law noise can all be seen.

Sometimes the shape of the observed power spectrum is ambiguous with respect
to the decomposition into power spectral components as described here. An
advantage of the $\chi^2$ minimization method is that it allows to quantify
the degree of this ambiguity by comparing the $\chi^2$ values of the different
possible combinations of fit parameters.

\section{Problems at the low-frequency end: \hfill
low-frequency leakage}\label{low_f}

At the low frequency end (typically, below 0.01\,Hz) of the power spectra a
number of problems occurs in an analysis along the lines described above that
is usually not really satisfactorily dealt with.

One problem is, that in order to reach the lowest frequencies in the first
place, it is necessary to choose the length of the time segments $T$
relatively large and therefore $M$ relatively small. This means that at the
lowest frequencies the statement made in Section\,\ref{ana}, that a large
number of individual powers has been averaged and therefore the average power
is approximately normally distributed is no longer true. Other (e.g.,
maximum-likelihood) fitting procedures are required to take the true
distribution of the average powers into account, but such methods are not
usually applied. See Papadakis and Lawrence (1993) for a discussion of a
method to remedy some of these problems.

A more serious problem is, that the true power spectrum at these frequencies
often seems to be a quite steep power law. The finite time window $T$ in those
situations leads to so-called ``low-frequency leakage'' (see Deeter 1984 and
references therein): power shows up at a higher frequency in the power
spectrum than where it belongs. One way to see this is by noting that the
lowest frequency accessible in the discrete Fourier transform is $1/T$. If
there is a lot of variability at lower frequencies than this, then due to
these slow variations the time series of an individual time segment will
usually have a large overall trend. The Fourier transform of this trend
produces a power law with index $-$2 in the power spectrum. Another way to
describe the effect of low-frequency leakage is by noting that according to
the Fourier convolution theorem the Fourier transform calculated in the finite
time window $T$ is related to the true Fourier transform by a convolution of
the true transform with the transform of the window function. As the window
function is a boxcar, its Fourier transform is the well-known sinc function,
with a big central lobe and upper and lower sidelobes that gradually decrease
in amplitude. For true power spectra steeper than a power law of index $-$2,
the contribution to the convolution of the lower side lobes overwhelms that of
the central lobe, and the result is a power spectrum that is a power law with
index $-$2 (e.g., Bracewell 1965). So, for any true power spectrum steeper
than $-$2, the actually measured power spectrum will have slope of
\about$-$2.

There are well-known solutions to this well-known problem. The most famous one
is data tapering: instead of a boxcar window a tapered window is used, i.e., a
window that makes the data go to zero near its end points more gradually than
by a sudden drop to zero. Other methods are polynomial detrending (fitting a
polynomial to the data and subtracting it) and end-matching (subtracting a
linear function that passes through the first and the last data point). Deeter
et al. (1982) and Deeter (1984) have explored a number of non-Fourier
methods. All these methods work in the sense that to some extent they suppress
the side lobes of the response function and therefore they are able to recover
power laws steeper than with index $-$2 (the value of the power law index
where the method breaks down is different in each case). However, typically
these methods have only been evaluated for the case where the time series is
pure power-law noise, and in many cases even only with respect to their
effectiveness in recovering the power law index, not even the noise
strength. Some methods require that the index of the power law is known in
advance! Nearly nothing is known about the way in which these methods affect
the results of fits to power spectra with complicated shapes such as those
described above. These methods may recover the correct power law index for the
low-frequency part of the power spectrum, but what will be the effect on the
fractional rms values and time scales of all the other components? For this
reason, these methods are not usually applied.

\section{Problems at the high frequency end: \hfill 
what is the shortest time scale?}\label{hi_f}

The high-frequency end of the power spectra, near the Nyquist frequency, is of
particular interest to astrophysicists, as it is there that we expect to find
the signatures of the fastest accessible physical processes going on in the
star. A question that is often asked in this context is: ``what is the
shortest variability time scale $\tau$ we can detect in the data?''.
Generally, what is observed at high frequency is that the power spectrum more
or less gradually slopes down towards the Poisson level. The problem is to
decide out to which frequency intrinsic power is observed above the Poisson
background, and {\it what this number means}.

An, I think, decidedly misleading way to answer the question about the
shortest time scale is by determining the {\it shortest time interval $\Delta
t$ within which significant variations can be detected}. It is obvious, that
for a slowly varying source, by zooming in on some gradual slope in its light
curve $I(t)$, this shortest time interval can be made as short as one wishes,
just by improving the quality of the data (by increasing $\epsilon$ for
example).  Yet, most of the work on ``shortest time scales'' seems to aim for
measuring $\Delta t$ rather than $\tau$. It seems clear that one must also
take into account the amplitude of the variations, not just the time within
which they occur to do something that is physically useful. Following Fabian
(1987) one can for example define the variability time scale as

$$\tau(t) = {I(t)\over \dot I(t)},$$

\noindent where the dot denotes the time derivative. 
Defined this way, $\tau$ is a measure of how steeply $I$ changes with $t$, and
depends itself on $t$. Using a power spectrum, one would measure some average
of this quantity by determining the fractional rms amplitude $r(\nu_{hi})$
near some high frequency $\nu_{hi}$ in the way described in
Section\,\ref{ana}, and then write

$$\overline\tau = C \times {1\over \nu_{hi}r(\nu_{hi})},$$

\noindent where $C$ is a constant of order unity depending on assumptions on 
exactly how the variations causing the power in the spectrum took
place. However, for precise work one has to worry about low-frequency leakage,
too. Exactly the same problem as described in Section\,\ref{low_f} can occur
here: power from lower frequencies can leak up to higher frequencies due to
the sinc response associated with the power estimators. That this problem is
serious is apparent from the fact that the observed power spectrum of for
example the famous black-hole candidate Cyg\,X-1 has an index of \about$-$2
for frequencies above \about10\,Hz (see Fig.\,\ref{bhcps}), just the value at
which low-frequency leakage begins to worry us. It would be of great interest
to have a foolproof way to subtract power that has leaked up from lower
frequencies, or even to have a way to make a conservative estimate of (obtain
a lower limit on) the true high-frequency power. It is not clear to what
extent this can be accomplished. Obviously, as in any convolution problem,
some information has been lost, but how much can be recovered is a problem
X-ray astronomers do not know the answer to. I wonder to what extent standard
deconvolution procedures might be useful here.

I note parenthically that a method that has been applied for determining the
shortest variability in a time series by Meekins et al. (1984) seems to
suffer from the same problem of low-frequency leakage. In this method, the
time series is divided up in very short segments of, for example, $N=10$
bins each. In each segment a quantity called ``chi-squared'' is calculated
as follows: 

$$\chi^2_{Meekins} = \sum_{k=0}^{N-1}{(x_k-N_{ph}/N)^2\over N_{ph}/N}.$$

Here $x_k$ is the number of photons detected in bin $k$ and $N_{ph}$ is the
total number of photons in the segment. One recognizes an ``observed over
expected'' variance ratio for an expected Poisson distribution. The
distribution of this quantity is then compared to that expected if all
variability in the time series would be due to Poisson fluctuations and if a
significant excess is found, this is interpreted as detection of variability
on time scales between the length $T$ of a segment and the duration of one
bin. One would expect low-frequency leakage to be as much of a problem here as
in the power spectral method. It is easy to see that if there are variations
in the time series on time scales much longer than the segment length, the
data points in the segments will usually follow steep trends, causing large
values of $\chi^2_{Meekins}$ that are not related to variability on time
scales shorter than $T$. Indeed, Meekins et al. show that $\chi^2_{Meekins}$
is closely related to the Fourier power in the segment, so from the point of
view of low-frequency leakage the Meekins et al. method is less effective than
the standard power spectral approach, as it requires $T$ to be quite short and
therefore increases the probability of steep trends in the data segments.

\bibliographystyle{alpha}

\end{document}